\documentclass[aps,prl,twocolumn,superscriptaddress]{revtex4}
\usepackage{amsmath}
\usepackage{graphicx}
\usepackage[section]{placeins}
\usepackage{subfigure}

\newcommand{\ket}[1]{|#1\rangle}
\newcommand{\bra}[1]{\langle#1|}

\newcommand{\eq}{\begin{equation}}
\newcommand{\fine}{\end{equation}}

\begin{document}

\title{Testing sequential quantum measurements: how can maximal knowledge be extracted?}
\author{Eleonora Nagali}

\affiliation{Dipartimento di Fisica, Sapienza Universit\`{a} di Roma, Roma 00185, Italy}

\author{Simone Felicetti}

\affiliation{Dipartimento di Fisica, Sapienza Universit\`{a} di Roma, Roma 00185, Italy}

\author{Pierre-Louis de Assis}

\affiliation{Departamento de Fisica, Universidade Federal de Minas Gerais,\\ Caixa Postal 702, 30123-980, Belo Horizonte, Brazil}

\author{Vincenzo D'Ambrosio}

\affiliation{Dipartimento di Fisica, Sapienza Universit\`{a} di Roma, Roma 00185, Italy}

\author{Radim Filip}

\affiliation{Department of Optics,
Palack\' y University, 17.~listopadu 12, 771~46 Olomouc, Czech Republic}

\author{Fabio Sciarrino}

\email{fabio.sciarrino@uniroma1.it}

\affiliation{Dipartimento di Fisica, Sapienza Universit\`{a} di Roma, Roma 00185, Italy}

\begin{abstract}
The extraction of information from a quantum system unavoidably implies a modification of the measured system itself. It has been demonstrated recently that partial measurements can be carried out in order to extract only a portion of the information encoded in a quantum system, at the cost of inducing a limited amount of disturbance. Here we analyze experimentally the dynamics of sequential partial measurements carried out on a quantum system, focusing on the trade-off between the maximal information extractable and the disturbance. In particular we consider two different regimes of measurement, demonstrating that, by exploiting an adaptive strategy, an optimal trade-off between the two quantities can be found, as observed in a single measurement process. Such experimental result, achieved for two sequential measurements, can be extended to $N$ measurement processes.
\end{abstract}
\maketitle

The measurement process represents one of the most distinctive aspects of quantum mechanics with respect to classical physics \cite{Von95,Brag95}. The main result of quantum measurement theory is the unavoidable disturbance of the quantum state by the measuring process, as epitomized by the early Heisenberg x-ray microscope thought experiment \cite{Heis30}. The duality between the information available on an unknown quantum system and the disturbance induced by a measurement process is of utmost relevance when investigating the quantum world \cite{Scul91,Bert01,Durr98} and lies at the basis of the security of quantum cryptographic protocols \cite{Gisi02}. In this framework, a measurement approach can be adopted to extract only a partial amount of information from the quantum system at the cost of limited induced decoherence \cite{Bana01,Acin05,Scia06,Ralp06}. Thus by adopting this partial measurement technique, consecutive observations (i.e. sequential measurements) can be carried out on the same quantum system to investigate its properties without destroying it \cite{Nogu99,Pryd04,Cabe08,Legg85,Amse09,Bart09,Kirch09}. The scenario of multiple measurement on the same quantum system leads to the following question: for a given amount of disturbance, how can knowledge from sequential measurements be optimally extracted and accumulated?

\begin{figure}[h!!!]
\centering
\subfigure{\includegraphics[scale=0.22,bb=0 0 800 600,clip]{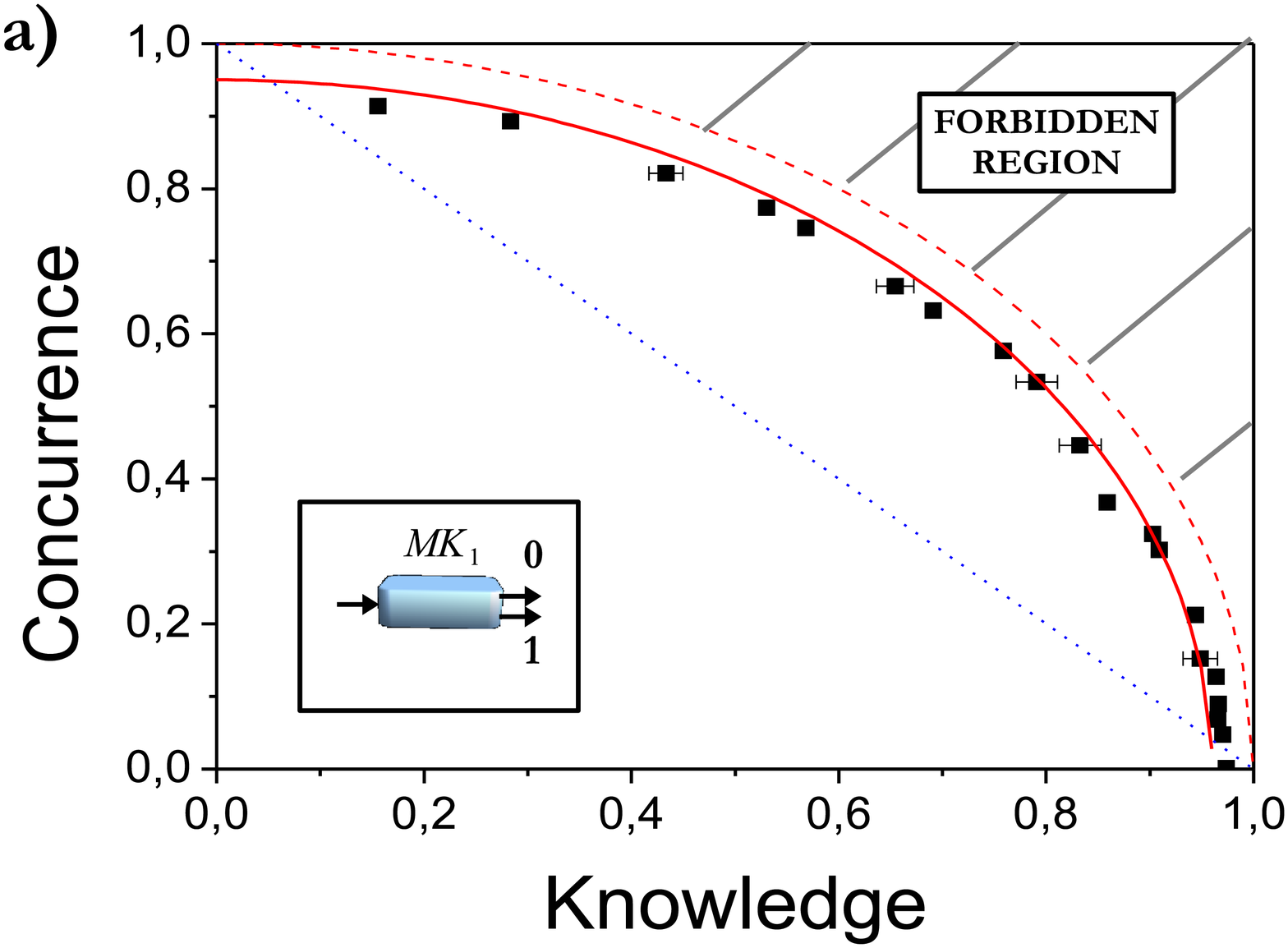}}
\caption{ \textbf{a)} Theoretical expectations compared to experimental data (dots) for concurrence  as function of the knowledge for the single measurement case. Red dashed line refers to a single measurement process, while continuous line takes into account the parameters of the $PBS$ ($t_H=r_V=0.992$,$r_H=t_V=0.008$) and the concurrence of the initial state. The blue dot line refers to the incoherent case without considering any imperfections of the system.. 
}
\label{kit1}
\end{figure}

The aim of this paper is to  investigate experimentally the trade-off between information gained and disturbance induced by partial sequential measurements on a quantum system adopting different measurement strategies \cite{Fili11}. In particular we compare the different trade-offs for two classes of measurement processes that can act on a quantum system: coherent and incoherent. Coherent measurements 
require that the system and measurement apparatus have a well defined phase 
relation (e.g. homodyne detection), while incoherent measurements are implemented as a projection that does not exploit the phase of the quantum state respectively to the measurement apparatus (e.g. photodetection). Such incoherent measurement process can be performed partially in the sense that the projection involves only a sub-ensemble of equally prepared quantum states, leaving the rest unmeasured. 
In this letter, we implement experimentally two sequential measurements performed on the same quantum system and observe that when sequential coherent measurements are carried out, the trade-off outperforms the one that characterizes incoherent sequential measurements. Moreover, the optimal trade-off that characterizes the single coherent measurement can be retrieved by adopting a proper adaptive strategy. Such result, observed for $N=2$ sequential measurements, can be extended for any value of $N$. \\
\begin{figure*}[t!!!]
\centering
\includegraphics[width=1\textwidth]{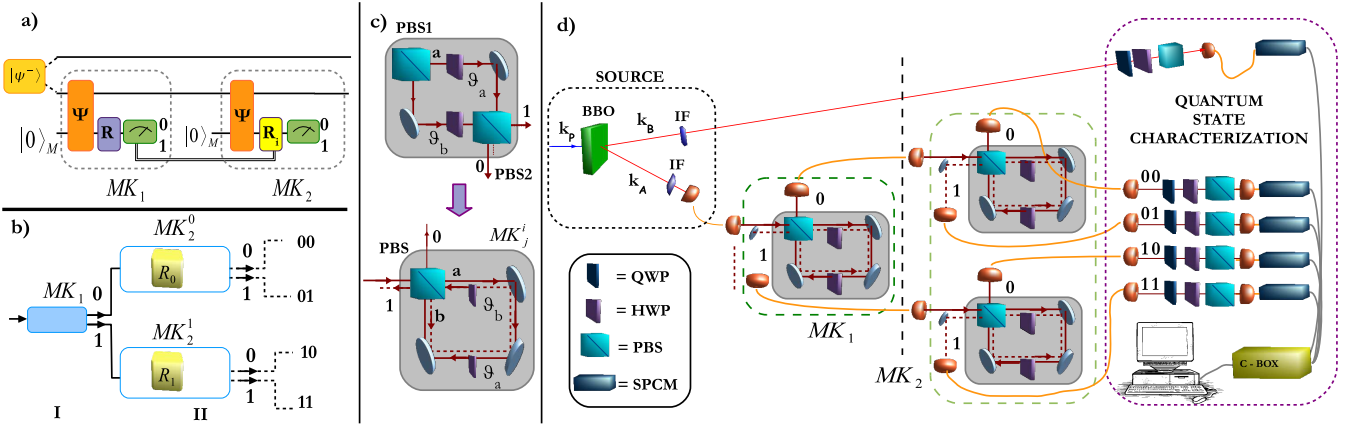}
\caption{ \textbf{a)} Scheme of the sequential measurements scenario. The qubit B is coupled with the ancillary qubit $\ket{0}_M$. The measurement result affects the next sequential one by adapting the measurement basis through a rotation $R_i$ in the next measurement process ($MK_2$). \textbf{b)} Scheme for the implementation of two sequential measurements strategy on single photon states.  An individual photon passing through the whole measurement apparatus is detected only in one of the four output ports, which indicates which 
combination of results for $MK_1$ and $MK_2$ is obtained. \textbf{c)}: Scheme of the measurement apparatus. Top: a polarizing Mach-Zehnder interferometer which allows to separately manipulate the horizontal and vertical polarizations adopting two waveplates oriented at angles $\theta_a$ and $\theta_b$, related by a shift of $\pi/4$ for the optimal configuration. The physical angle $\theta_b$ is related to the strength parameter $\psi$ by the relation $\psi=\frac{\pi}{4}-2\theta_b$, leading to $K=|cos(4\theta_b)| $.  
Bottom: to obtain a stable apparatus, an equivalent Sagnac interferometer has been adopted. \textbf{d)}: Experimental setup adopted for the sequential measurement strategy for $N=2$ measurement processes.}
\label{setup}
\end{figure*} 
The experimental investigation on the duality between decoherence and knowledge has been carried out through the interaction between one part of an entangled quantum system and an ancillary qubit (the `meter') on which projective measurements have been performed. This has also allowed us to observe a Zeno-like behavior \cite{Itan90} of the entanglement dynamics as a function of the strength of the interaction.\\
Let us define the figures of merit that describe the involved quantities:\\
i) The amount of information that can be gained on a quantum state is quantified by the \textit{knowledge} $K\in[0,1]$. Such parameter is the capability of correctly discriminating the quantum states belonging to the computational basis $\{\ket{0},\ket{1}\}$. Specifically $K$ is defined as  $K=|\sum_{i}p(i,i)-\sum_{i\neq j}p(i,j)|$ where $p(i,j)$ is the probability of guessing the state $\ket{i}$ when the input one is $\ket{j}$. $K=1$ corresponds to maximal knowledge and is achieved by projective measurements $\{\ket{0}\bra{0},\ket{1}\bra{1}\}$, while $K<1$ can be achieved by a partial measurement. When $K\rightarrow 0$ we perform a weak measurement. 
\\
ii) The disturbance effect of the measurement process on the initial maximally entangled state has been estimated by analyzing the degree of entanglement of the system, quantified by the concurrence $C\in[0,1]$, which gets lower ($C<1$)  as the information extracted from the system increases \cite{Woot98, Engl00}. \\
Let us first analyze the single coherent measurement strategy, represented by the quantum circuit in Fig.(\ref{setup}-\textbf{a}): box $MK_1$.  We consider a singlet state shared by parties $A$ and $B$ $\ket{\psi^-}_{AB}=\frac{1}{\sqrt{2}}(\ket{10}_{AB} -\ket{01}_{AB})$ generated by a standard entangled state source. The measurement strategy gains information on qubit B by entangling it with the meter $M$ through a coupling of variable strength, parametrized by $\psi$, and then performing a projective measurement on $M$. \\
Specifically, the qubit B interacts with $M$, initialized as $\ket{0}_M$, through the following unitary transformation: $\hat{U}(\psi)\ket{i}_B\ket{0}_M=\ket{i}_B\ket{\alpha_i}_M$ where $i=0,1$, and $\ket{\alpha_i}$ are two pure states.  Both $\ket{\alpha_i}$ can be expressed in terms of $\psi$: $\ket{\alpha_0}_M=\cos\psi\ket{0}_M+\sin\psi\ket{1}_M \quad ; \quad \ket{\alpha_1}_M=\cos\psi\ket{0}_M-\sin\psi\ket{1}_M$.
 In order to extract information from the system the meter is measured in the diagonal basis $\ket{\pm}_M=\frac{\ket{0}_M\pm\ket{1}_M}{\sqrt{2}}$, so that the optimal value of the knowledge, expressed following the definition in (i), is found to be  $K=|\sin2\psi|$. Hence it emerges that for the optimal coherent measurement described above, the residual entanglement is related to the knowledge extracted as $ C^{coh}=\sqrt{1-K^2}$ \cite{Engl00}, so that when all the information available is extracted ($K=1$), the initial entanglement is completely lost $(C=0)$.
Such behavior can be compared to that of incoherent measurements, which gives $C^{inc}=1-K$ \cite{Fili11}. \\
The former relations have been experimentally implemented generating two photons entangled in polarization, $\frac{1}{\sqrt{2}}(\ket{HV}_{AB} -\ket{VH}_{AB})$, where $(H,V)$ are linear horizontal and vertical polarization, respectively. Then, to carry out a single coherent measurement, we have exploited a different
degree of freedom of the same photon, encoding the meter qubit $M$ in the linear momentum, since this approach provides a natural phase relation between qubit B and the meter $M$.
For the measurement setup, hereafter denoted as \textit{measurement kit} $MK$, we refer to the schematic representation in Fig.(\ref{setup}-\textbf{c}).  The $MK$ corresponds to a Sagnac interferometer with a polarizing beam splitter ($PBS$), that interfaces the polarization to the output spatial modes $a$ (transmitted) and $b$ (reflected) \cite{Walb06}. On both internal propagation modes $a$ and $b$ of the Sagnac two half-waveplates, rotated at angles $\theta_a$ and $\theta_b$, allow to manipulate the polarization degree of freedom of the state. The selection of one of the output modes of the $PBS$ thus represents the analysis of the outcomes of the projection on the meter qubit. The knowledge $K$ has been estimated as $K=\frac{p(0,H)+p(1,V)-p(1,H)-p(0,V)}{2}$, from the definition (i), 
where $p(i,j)$ is the probability that an input photon with polarization $j$ will emerge on the mode corresponding to the outcome $i$. On the other hand the concurrence  of the state $\rho_{AB}$ after the single measurement process acting on qubit $B$ has been estimated from the density matrix reconstructed via quantum state tomography \cite{Woot98}. Experimental results are reported in Fig.(\ref{kit1}-\textbf{a}), and compared to theoretical expectations evaluated taking into account imperfections due to the $PBS$ and to the source of entangled states. \\
\begin{figure*}[t]
\centering
\includegraphics[scale=0.24]{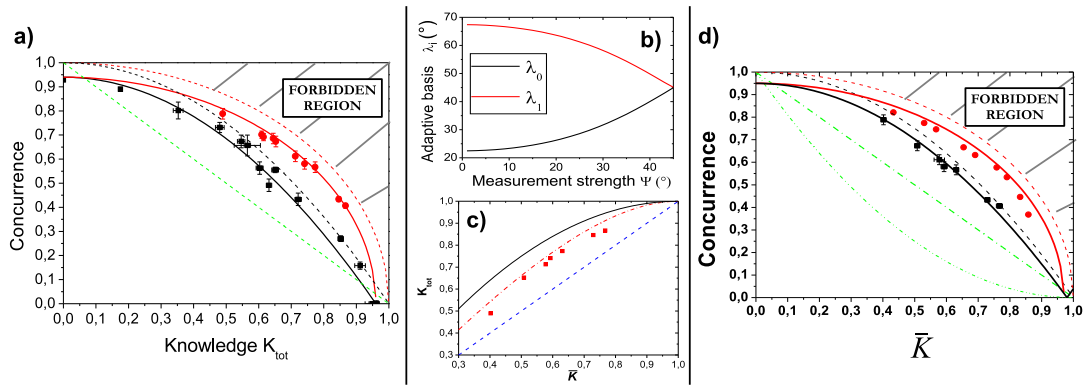}
\caption{\textbf{a)} Concurrence $C$ as function of the knowledge for $N=2$ sequential measurements adopting an independent strategy (black squares) and the adaptive one (red circles). Black and red lines represent theoretical expectations (dashed lines for the ideal case, continuous ones rescaled by experimental imperfections) for the two approaches to be compared to incoherent behavior (dashed green line). \textbf{b)} Numerical determination of adaptive basis depending on the measurement carried out in the first kit, expressed by the parameter $\psi$. \textbf{c)} Experimental knowledge after $N=2$ sequential adaptive measurements (red squares) compared to theoretical predictions for incoherent (black line), adaptive extraction (dashed-dot red line), and after $N=1$ measurement (dashed blue line). \textbf{d)} Experimental and theoretical behavior of concurrence as function of $\bar{K}$. Black squares and line refer to experimental $N=2$ adaptive measurements and theoretical expectations, respectively. Analogously red dots and line refers to the experimental and theoretical results for the single quantum measurements. The convexity of such behaviors are compared to the incoherent one, reported as green dashed-dot line ($N=1$) and green dashed-dot-dot line ($N=2$).}
\label{kit2}
\end{figure*}
Let us now address the following question: how can knowledge be extracted from sequential measurements?
We consider two sequential measurement processes of equal strength, represented by the schemes in Fig.(\ref{setup}-\textbf{a,b}). Each measurement process introduces a specific amount of decoherence, reflected by a lowering of the concurrence after each step. The degree of entanglement of the state after the two sequential measurements gives an indication of the total disturbance induced in the whole process.

For two subsequent incoherent measurements the concurrence scales linearly with the total amount of information $K_{tot}^{inc}$ extracted from the system $C^{inc}=1-K_{tot}^{inc}$: see Fig.(\ref{kit2}-\textbf{a}). 

As second benchmark, we consider two independent coherent sequential measurements, that is, the second projection on the state is performed independently of the outcome of the first one. In this case the concurrence shows a different dependence from the whole amount of knowledge extracted: $C^{ind}=1-(K_{tot}^{ind})^2$.  In Fig.(\ref{kit2}-\textbf{a}) we report the experimental behavior of the concurrence for this measurement strategy (black squares), where the total knowledge $K_{tot}$ has been evaluated as  in the single measurement procedure, combining outcomes $00$ with $01$ and $10$ with $11$. 
Comparing to the incoherent case, we observe a better trade-off between information extracted and decoherence introduced, even if the relation between the two parameters is not optimal.

The optimal trade-off between total decoherence and knowledge extracted from a quantum system is given by the relation $C^{opt}=\sqrt{1-K_{tot}^2}$, analogously to what observed for the single coherent measurement. In order to achieve such trade-off  it has been theoretically shown \cite{Fili11} that an adaptive strategy can be adopted, similar to the one proposed  for discrimination of multiple copies of quantum states \cite{Acin05}. As schematically shown in Fig.(\ref{setup}-\textbf{a}), the results from the first measurement kit determine an adaptation, i.e. a rotation in the meter basis, for the subsequent measurement process.
Therefore, classical communication is required between the sequential measurements and they cannot be treated as independent anymore.
Depending on the outcome $0$ or $1$ of $MK_1$, two different basis of analysis, generically indicated as $\{\ket{\beta_0},\ket{\beta_0^{\perp}}\}$ and $\{\ket{\beta_1},\ket{\beta_1^{\perp}}\}$, are applied on the meter qubit in $MK_2^0$ and $MK_2^1$, where $ \ket{\beta_i}=\cos\lambda_i\ket{0}+\sin\lambda_i\ket{1}\nonumber $. Both parameters $\{\lambda_0,\lambda_1\}$ are determined in order to maximize the extracted knowledge and depend from the decoherence induced by the first measurement process.  In Fig.(\ref{kit2}-\textbf{b}) we report the numerical determination of parameters $\{\lambda_0,\lambda_1\}$ depending on the measurement strength $\psi$ of the first kit. \\
The adaptive strategy has been implemented experimentally by rotating the waveplates in the Sagnac of $MK_2$, thus modifying the basis of the meter qubit depending on the measurement carried out by $MK_1$. We note that an intrinsic feed-forward takes place in the adaptation process since two different rotations of the basis are performed in the second measurement process, depending on which output arm of the first interferometer the photon gets out, as shown in Fig. (\ref{setup}-\textbf{b,d}). In Fig.(\ref{kit2}-\textbf{a}) we report the theoretical behavior of concurrence as function of the global knowledge $K_{tot}$ and the experimental results. The value of $K_{tot}$ has been estimated with the same relation adopted for the single measurement process, where outcomes $i$ refer only to the second kit. 
We find a good agreement with theoretical predictions rescaled to experimental imperfections. We conclude that by performing two sequential measurements with the adaptive strategy, we find the same optimal trade-off that characterizes the single coherent measurement. \\
In Fig.(\ref{kit2}-\textbf{c}) we compare how knowledge accumulates after $N=2$ incoherent (continuous black line) and adaptive coherent measurements (red squares and dashed-dot line) as a function of $\bar{K}$, the knowledge that would be extracted from each measurement if it were the only one performed. As the total amount of knowledge extractable is equal to $1$, after the first measurement there is less knowledge available to be extracted by the second process. Therefore, as shown in \cite{Fili11}, the total knowledge extracted after any two sequential measurements is not simply $\bar{K_1}+\bar{K_2}$. In particular we observe that, in the case of two measurements of equal $\bar{K}$, more knowledge is accumulated for the incoherent case compared to the adaptive one. Considering both Fig.(\ref{kit2}-\textbf{c}) and Fig.(\ref{kit2}-\textbf{d}) we see that although the incoherent strategy accumulates more knowledge, it lowers the concurrence at a much faster rate thus confirming the advantage in adopting a coherent adaptive strategy.
Finally, Fig.(\ref{kit2}-\textbf{d}) experimentally demonstrates that the adaptive concurrence $C^{adapt}_{tot}$ after two identical measurements is a concave function of $\bar{K}$, as in the single measurement case. On the other hand, for incoherent measurements the function $C^{inc}(\bar{K})$ appears to be convex  both for $N=1$ and $N=2$. This qualitative distinction for the curvature of the function is responsible for a Zeno-like effect observed for coherent measurements \cite{Fili11}. Indeed the quantum Zeno-effect refers to a situation in which a quantum system, if observed frequently by projective measurements, varies slower than the exponential decay law. Here the
concurrence scales with $\bar{K}$ as $C^{adapt}_{tot}(\bar{K})\approx 1-N\frac{\bar{K}^2}{2}$, while for incoherent
measurements we get $C(\bar{K})\approx 1-N\bar{K}$.
Thus for a given $\bar{K}$ the coherent measurements show a concurrence which degrades slower than in the incoherent case.  Moreover in the coherent scenario, the amount of entanglement is
weakly affected when a small amount of information is acquired in each single measurement. This Zeno-like
behavior is a consequence of the concave dependence of the concurrence from the knowledge $\bar{K}$  after any $N$
sequential measurement in the
coherent case, to be compared to a strong convex dependency for the incoherent ones. The coherence
in the measuring apparatus thus becomes a basic resource for quantum controlled dynamics.\\
In summary, we have reported the experimental analysis on the trade-off between acquired knowledge from a quantum state and the detrimental effect on the system itself, performing sequential coherent measurements. We have experimentally investigated how knowledge can be accumulated from two sequential coherent 
measurements, observing that an optimal trade-off can be achieved when an adaptive strategy  is adopted. Finally we have observed  the fundamental influence of the coherence on knowledge accumulation for quantum measurements, which can be exploited for deterministic Zeno-like behavior. Future steps might be the study of the extension for multiple measurements and the application to quantum error correction techniques and others quantum information protocols \cite{Yoko09,Giov10}.\\
This work was supported by project FIRB- Futuro in Ricerca (HYTEQ),
Finanziamento Ateneo 2009 of Sapienza Universita` di
Roma, and European project PHORBITECH of the FET
program (Grant No. 255914). P.-L. de Assis acknowledges the financial support of the CNPq, CAPES and FAPEMIG Brazilian research funding agencies.
R.F. acknowledges grants MSM6198959213, LC06007 and ME10156 of MSMT CR.

%

\begin{thebibliography}{10}

\bibitem{Von95}
J.~Von~Neumann,
\newblock {\em {Mathematische grundlagen der quantenmechanik}} (Springer,
  1995).

\bibitem{Brag95}
V.~Braginsky, F.~Khalili, and K.~Thorne,
\newblock {\em {Quantum measurement}} (Cambridge Univ Pr, 1995).

\bibitem{Heis30}
W.~Heisenberg and C.~Eckart,
\newblock {\em {The physical principles of the quantum theory}} (Dover Pubns,
  1930).

\bibitem{Scul91}
M.~Scully, B.~Englert, and H.~Walther,
\newblock Nature {\bf 351}, 111 (1991).

\bibitem{Bert01}
P.~Bertet {\em et~al.},
\newblock Nature {\bf 411}, 166 (2001).

\bibitem{Durr98}
S.~Durr, T.~Nonn, and G.~Rempe,
\newblock Nature {\bf 395}, 33 (1998).

\bibitem{Gisi02}
N.~Gisin, G.~Ribordy, W.~Tittel, and H.~Zbinden,
\newblock Rev. Mod. Phys. {\bf 74}, 145 (2002).

\bibitem{Bana01}
K.~Banaszek,
\newblock Phys. Rev. Lett. {\bf 86}, 1366 (2001).

\bibitem{Acin05}
A.~Acin, E.~Bagan, M.~Baig, L.~Masanes, and R.~Munoz-Tapia,
\newblock Phys. Rev. A {\bf 71}, 032338 (2005).

\bibitem{Scia06}
F.~Sciarrino, M.~Ricci, F.~De~Martini, R.~Filip, and L.~Mi{\v{s}}ta~Jr,
\newblock Phys. Rev. Lett. {\bf 96}, 20408 (2006).

\bibitem{Ralp06}
T.~Ralph, S.~Bartlett, J.~O�Brien, G.~Pryde, and H.~Wiseman,
\newblock Physical Review A {\bf 73}, 12113 (2006).

\bibitem{Nogu99}
G.~Nogues {\em et~al.},
\newblock Nature {\bf 400}, 239 (1999).

\bibitem{Pryd04}
G.~J. Pryde, J.~L. O'Brien, A.~G. White, S.~D. Bartlett, and T.~C. Ralph,
\newblock Phys. Rev. Lett. {\bf 92}, 190402 (2004).

\bibitem{Cabe08}
A.~Cabello,
\newblock Phys. Rev. Lett. {\bf 101}, 210401 (2008).

\bibitem{Legg85}
A.~J. Leggett and A.~Garg,
\newblock Phys. Rev. Lett. {\bf 54}, 857 (1985).

\bibitem{Amse09}
E.~Amselem, M.~Radmark, M.~Bourennane, and A.~Cabello,
\newblock Phys. Rev. Lett. {\bf 103}, 160405 (2009).

\bibitem{Bart09}
H.~Bartosik, J.~Klepp, C.~Schmitzer, S.~Sponar, and Cabello,
\newblock Phys. Rev. Lett. {\bf 103}, 4 (2009).

\bibitem{Kirch09}
G.~Kirchmair {\em et~al.},
\newblock Nature {\bf 460}, 494 (2009).

\bibitem{Fili11}
R.~Filip,
\newblock Phys. Rev. A {\bf 83}, 032311 (2011).

\bibitem{Itan90}
W.~Itano, D.~Heinzen, J.~Bollinger, and D.~Wineland,
\newblock Physical Review A {\bf 41}, 2295 (1990).

\bibitem{Woot98}
W.~K. Wootters,
\newblock Phys. Rev. Lett. {\bf 80}, 2245 (1998).

\bibitem{Engl00}
B.~Englert and J.~Bergou,
\newblock Opt. Comm. {\bf 179}, 337 (2000).

\bibitem{Walb06}
S.~Walborn, P.~Ribeiro, L.~Davidovich, F.~Mintert, and A.~Buchleitner,
\newblock Nature {\bf 440}, 1022 (2006).

\bibitem{Yoko09}
K.~Yokota, T.~Yamamoto, M.~Koashi, and N.~Imoto,
\newblock New Journal of Physics {\bf 11}, 033011 (2009).

\bibitem{Giov10}
V.~Giovannetti, S.~Lloyd, and L.~Maccone,
\newblock Arxiv preprint arXiv:1012.0386  (2010).

\end{thebibliography}

\end{document}